\documentclass[lettersize,journal]{IEEEtran}
\usepackage{amsmath,amsfonts}
\usepackage{algorithmic}
\usepackage{algorithm}
\usepackage{array}
\usepackage{color}
\usepackage[caption=false,font=normalsize,labelfont=sf,textfont=sf]{subfig}
\usepackage{textcomp}
\usepackage{stfloats}
\usepackage{url}
\usepackage{verbatim}
\usepackage{graphicx}
\usepackage{cite}
\usepackage{booktabs} 
\usepackage[colorlinks=true,
            citecolor=blue,
            linkcolor=blue,
            urlcolor=blue]{hyperref}

\usepackage{url}
\allowdisplaybreaks[4]

\begin{document}
\title{LMMSE-Based Phase Estimator for Fiber-Optic Coherent Spread-Spectrum FBG-Based Sensing Systems}

\author{Yiyang Bai, \'Elie Awwad, and Mich\`ele Wigger%~\IEEEmembership{Staff,~IEEE,}
        % <-this % stops a space
\thanks{Y. Bai and E. Awwad are with Télécom Paris, LTCI, IP Paris. Email: \{yiyang.bai, elie.awwad\}@telecom-paris.fr.  M. Wigger is with CentraleSup\'elec, L2S, Universti\'e Paris-Saclay. Email: michele.wigger@centralesupelec.fr}}% <-this % stops a space
%\thanks{Manuscript received April 19, 2021; revised August 16, 2021.}}

% The paper headers
%\markboth{Journal of \LaTeX\ Class Files,~Vol.~14, No.~8, August~2021}%
%{Shell \MakeLowercase{\textit{et al.}}: A Sample Article Using IEEEtran.cls for IEEE Journals}

%\IEEEpubid{0000--0000/00\$00.00~\copyright~2021 IEEE}
% Remember, if you use this you must call \IEEEpubidadjcol in the second
% column for its text to clear the IEEEpubid mark.

\maketitle

\begin{abstract}
Fiber Bragg grating (FBG)-based sensing enables localized strain and temperature monitoring by tracking the phase changes of the light reflected from FBGs inserted along a fiber. A major source of disturbance in such systems is laser phase noise, and its influence is typically mitigated by means of additional hardware. In this article, we propose to mitigate the influence of laser phase noise by means of digital signal processing. We propose a practical estimator for coherent single-polarization spread-spectrum FBG-based sensing systems, which is motivated by linearizing the complex received signal and approximating the loss function to formulate a minimum mean square error (MMSE) estimation problem. Moreover, a Cramér–Rao lower bound for this FBG-based estimation problem is presented. Our simulation results demonstrate that the performance of our proposed estimator lies close to the Cramér–Rao lower bound, thus indicating that our estimator effectively compensates for  phase noise in  FBG-based sensing systems.
\end{abstract}


\section{Introduction}
\IEEEPARstart{F}{iber} Bragg grating (FBG)-based optical fiber sensing has emerged as a promising tool for internal and external temperature monitoring of lithium-ion batteries~\cite{han2021review}, structural health monitoring of bridges and buildings~\cite{majumder2008fibre}, and non-contact force sensing for micro/nano manipulation systems~\cite{zang2019recent}. Indeed, FBG sensing systems can provide high sensing accuracy while causing little interference and remaining robust under various harsh environmental conditions. By embedding FBGs with distinct Bragg wavelengths along the fiber, the wavelength-selective reflection of each grating enables sensing at specific locations, as the phase shift of the signal reflected from each FBG captures local perturbations. Among various FBG-based sensing techniques, spread-spectrum probing~\cite{Dor_18} continuously transmits low-power, encoded waveforms into the fiber and recovers spatial information through a correlation process. Unlike pulse-based~\cite{pastor2016single} or chirped-pulse~\cite{zou2015optical} probing, its constant-power operation avoids high peak powers, thereby significantly reducing nonlinear interference when applied on a lit-fiber in parallel to data-modulated wavelengths in the same transmission band.

A common probing technique in FBG-based sensing is phase-sensitive optical time domain reflectometry ($\phi$-OTDR), which detects external perturbations by injecting coherent signals and analyzing the interference phase of backscattered light~\cite{zhang2015new}. A major challenge in these systems is the phase noise of the laser, which severely contaminates the phase of the sensing signal\cite{modeling} and degrades sensing accuracy %. Therefore, laser phase noise is generally recognized as a key limiting factor that degrades accurate sensing parameter measurements
\cite{Shujie2015Influence}.% in FBG-based sensing systems. 

Previous works have proposed hardware-based solutions to suppress phase noise at the physical source or through a specialized system design, for example using advanced lasers. Since phase noise is largely related to the laser's linewidth, employing ultra-narrow linewidth lasers or implementing a stabilization technique for such lasers  reduces the noise at their source\cite{laser}. However, such high-performance laser solutions are  costly and hinder large-scale deployment. Another strategy is to measure the phase noise in the FBG-based sensing systems using additional hardware. For example, an auxiliary interferometer can be added to monitor the laser's instantaneous frequency drift in real-time\cite{Liu2025UWFBG}. This measured noise can then be subtracted from the main sensing signal at the receiver for compensation. While effective in mitigating phase noise at specific locations in the fiber, this method introduces additional hardware complexity and cost, which again limits its large-scale applicability.

Instead of hardware solutions, a few studies suggested to employ digital signal processing methods for phase noise mitigation. Prominent data-driven approaches utilize deep learning to filter noise directly from the measured FBG responses\cite{tiwari2024deep,Liu2025Fiber}. Yet, their effectiveness heavily relies on extensive training datasets and lack theoretical interpretability.
Alternatively, we adopt a model‑driven methodology, similar to the one suggested by Yang et al.~\cite{yang2009near} for phase estimation in a wireless communication system. In their work, a near‑optimal phase estimator is derived from the maximum a posteriori (MAP) framework to mitigate phase noise in a wireless communication scenario, in which phase noise is generated from two independent oscillators. Hence, the total phase noise is the sum of two independent noise processes added at the transmitter and the receiver respectively. In contrast, in coherent FBG-based sensing systems, the transmitter and the receiver are on the same side of the optical fiber and share the same laser source. In FBG-based systems, we generate a probing signal from a laser source, transmit it into the fiber, then detect the backscattered parts of this signal using the same laser. This leads to a differential phase noise term. In this work, we design a new estimator tailored to FBG-based sensing systems, with the goal of mitigating the impact of this phase-noise term. Our focus is therefore on the statistical estimation error, referred to as Type‑A uncertainty. For a broader discussion, also involving Type‑B uncertainty, see \cite{B-type}.
%Our focus is therefore on the statistical estimation error. For a broader discussion, involving both Type-A and Type-B uncertainties, see \cite{B-type}}.

\IEEEpubidadjcol
Specifically, we propose a practical estimator for spread-spectrum single-reflector FBG sensing systems, and prove efficiency of this estimator through numerical simulations, in which the mean square error (MSE)  of the proposed estimator is compared with that of a standard matched filter estimator and with the Cram\'er-Rao lower bound (CRLB). 
%on the sensing limit e linear minimum mean square error (LMMSE)-based estimator to achieve accurate sensing.
The detailed contributions of this paper are:
\begin{itemize}
    \item A discrete-time signal model for the fiber-optic spread-spectrum single-polarization FBG-based sensing system, by modeling the phase noise as a temporally correlated Wiener process. We also introduce a circular loss function to measure accuracy of phase estimators.
    \item An approximation of our proposed signal model and loss function to obtain a classic Gaussian estimation problem with a squared-error loss, yielding a practical estimator based on this linear model.
    \item A derivation of CRLB for our original signal model and a numerical evaluation based on Monte-Carlo simulation. A simplified CRLB bound that ignores the additive noise component is also presented. It lower bounds the original CRLB, but is numerically more stable than the original bound at high signal-to-noise ratios (SNR).
    \item Numerical simulation results showing that our proposed estimator, when evaluated on the original signal model and loss function, outperforms the simple estimator and performs close to the CRLBs. This indicates that our proposed simple practical estimator mitigates the phase noise in a close to optimal way.
\end{itemize}

The paper is organized as follows. In Section II, we present the system model and the estimation problem, and %discrete-time model for the fiber-optic spread-spectrum single-polarization FBG-based sensing system and the estimation problem. 
in Section III, we introduce our practical estimator. Section IV provides the CRLB and Section V presents simulation results. Finally, Section VI concludes the paper and discusses future research.

\section{System Model and Estimation Problem}
\subsection{System Architecture and Continuous-Time Model}
\begin{figure}[!ht]
    \centering
    \includegraphics[width=0.8\linewidth]{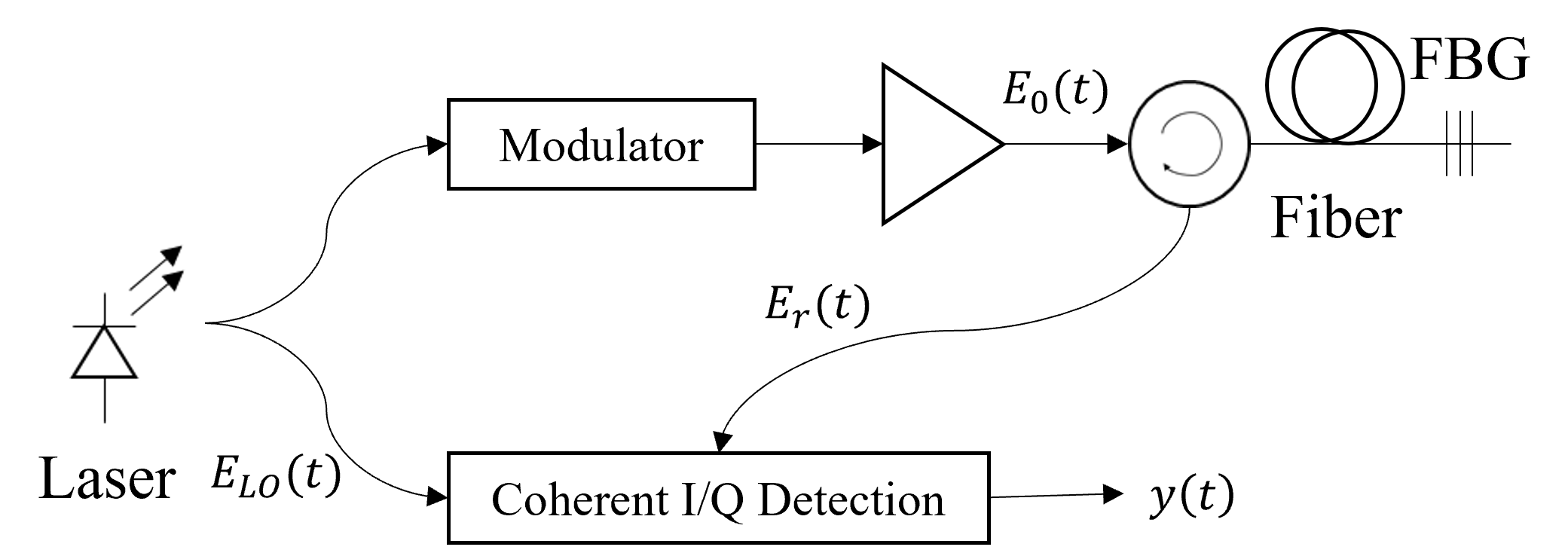}
    \caption{Schematic diagram of the spread-spectrum FBG-based sensing system.}
    \label{fig:system_architecture}
\end{figure}

Our sensing system is schematically illustrated in Fig.~\ref{fig:system_architecture}. 

The system employs a single continuous-wave laser source whose output is split into two paths: the transmitter arm and the local oscillator (LO) arm. In the transmitter arm, the optical carrier is modulated by a known complex-valued signal $s(t)$ %, such as a Constant Amplitude Zero Autocorrelation (CAZAC) sequence\cite{Rohrs1992CAZAC}, 
to generate the transmitted optical signal:
\begin{equation}
E_{0}(t)=A_{0}  s(t) \cdot e^{j\left[\omega_{c}t+\theta_{r}(t)\right]},
\label{eq:transmitted_field}
\end{equation}
where $A_0$ represents the transmit amplitude, $\omega_c$  the optical carrier angular frequency, and $\theta_{r}(t)$  the laser phase noise. %This modulated signal $E_0(t)$ is launched into the sensing fiber.

%After a round-trip propagation delay $\tau$ through the fiber, the backscattered signal from the sensing region arrives at the receiver. Actually, 
The sensing receiver observes the reflected signal from the FBG in the fiber. In this case, the received optical signal can be expressed as:
\begin{equation}
E_{r}(t)=A_{r} s(t-\tau(t)) \cdot e^{j\left[\omega_{c}(t-\tau(t))+\gamma(t)+\theta_{r}(t-\tau(t))\right]},
\label{eq:received_field}
\end{equation}
where $ \tau(t)=\tau+\Delta\tau(t)$ is the round-trip time, with $\tau$ denoting the static round-trip delay in the absence of any strain before the reflector, and $\Delta\tau(t)$ the additional delay induced by strain accumulated along the fiber before the reflector. The term $\gamma(t)$ denotes the phase change directly induced by the reflector. The amplitude $A_r$ accounts for attenuation due to scattering and fiber loss, and $\theta_{r}(t-\tau(t))$ is the delayed laser phase noise. 

%\yb{In typical FBG-based sensing systems, the strain-induced delay variation $\Delta\tau(t)$ is on the order of picoseconds or less, which is far smaller than the symbol period $T_s$, which we introduce when passing to the discrete and typically lies in the order of nanoseconds. Therefore, we can safely approximate $\tau(t) \approx \tau$ in the slowly varying envelope $s(t-\tau(t))$ and in the phase noise term $\theta_r(t-\tau(t))$. This yields:}
%\begin{equation}
%E_{r}(t) \approx A_{r} s(t-\tau) \cdot e^{j\left[\omega_{c}(t-\tau(t))+\gamma(t)+\theta_{r}(t-\tau)\right]}.
%\label{eq:received_field_approx}
%\end{equation}

The sensing receiver also has access to the local oscillator (LO) signal, derived from the same laser source as the transmit signal. It is described by:
\begin{equation}
E_{LO}(t)=A_{LO} \cdot e^{j\left[\omega_{c}t+\theta_{r}(t)\right]}.
\label{eq:lo_field}
\end{equation}
where $A_{LO}$ is the amplitude of the LO. The LO shares the same instantaneous phase noise $\theta_{r}(t)$ as the transmit signal, which is a crucial feature of this coherent architecture.

The sensing receiver feeds the  signals $E_r(t)$ and $E_{LO}(t)$  into an optical $90^\circ$ hybrid circuit for coherent I/Q detection\cite{Kikuchi2016Fundamentals}.  A perfect polarization alignment is assumed, which means that when $E_{LO}(t)$ and $E_{r}(t)$ reach the coherent I/Q detection, their polarization states are identical. As a result, we can simplify the complex baseband signal to:
\begin{eqnarray}
y(t) \! = \! As(t-\tau(t)) e^{j[-\omega_{c}\tau(t)+ \theta_d + \gamma(t) + \theta_r(t-\tau(t))-\theta_r(t)]} \hspace{-1cm}\nonumber \\
+b(t),\hspace{5cm}
\label{eq:baseband_signal}
\end{eqnarray}
where $A=2A_r A_{LO}$, % the term $-\omega_c\tau(t)$ represents the phase offset due to the fiber delay, $\omega_c\Delta\tau(t)$ is the dynamic phase contribution from strain before the reflector, 
$\theta_d$ is a constant phase difference caused by a constant optical-path difference between $E_r(t)$ and $E_{LO}(t)$, the inputs of the coherent receiver, and $b(t)$ is an additive white Gaussian noise (AWGN).

\subsection{Discrete-Time Signal Model}
The continuous-time signal $y(t)$ is sampled at the symbol period $T_s$, over $M$ blocks each consisting of $N$ symbols. In typical FBG-based sensing systems, the strain-induced delay variation $\Delta\tau(t)$ is in the order of picoseconds or less, which is far smaller than the symbol period $T_s$, which lies in the order of nanoseconds. Therefore, we can safely approximate $\tau(t) \approx \tau$ in the slowly varying envelope $s(t-\tau(t))$ and in the phase noise term $\theta_r(t-\tau(t))$. In addition, we assume that the strain-induced delay $\Delta\tau(t)$ and the reflector phase $\gamma(t)$ are constant over each block of duration $N T_s$. They both vary with the external perturbation, and their rate of change is thus limited by the maximum frequency $f_{\max}$ of this perturbation. Assuming that we choose a symbol period $T_s$ such that the duration of the entire block $NT_s$ is much smaller than the period of this  perturbation $N T_s \ll 1/(2f_{\max})$,\footnote{In practical systems $N$ is typically in the order of hundreds and $T_s$ in the order of nanoseconds. The assumption $N T_s \ll 1/(2f_{\max})$ thus is valid for all frequencies in the kHz regime, covering most acoustic and seismic signals.}  %ensuring that $\gamma(t)$ remains essentially constant over each block.
 $\Delta\tau(t)$ and $\gamma(t)$ remain almost constant over a block of $N$ symbols.

We thus denote the $m$-th block reflector phase by $\gamma_m$, and the delay variation by $\Delta\tau_m$, for $m=1,\ldots, M$. We further approximate the static delay as an integer multiple of the symbol period, i.e., $\tau = \ell T_s$ for a given  integer $\ell>0$, and assume that $s(t)$ is a periodic signal with period $NT_s$. Notice that here the delay parameter $\ell$ is proportional to the distance between the probing interrogator and the FBG.
We then obtain the discrete-time signal for the $n$-th symbol in the  $m$-th block:
\begin{IEEEeqnarray}{rCl}
y^{(m)}_{n} &=& A s_{n} \cdot e^{j(\psi_{m} + \xi^{(m)}_{n})} + b^{(m)}_{n}, \quad n = 1, \dots, N,\nonumber\\
&& \hspace{4.1cm}m=1,\ldots, M,
\label{eq:discrete_signal}
\end{IEEEeqnarray}
where the parameters are defined as:
\begin{itemize}
    \item $A=2A_r A_{LO}$;
    \item $s_n=s((n-\ell)T_s)$;
   \item $\psi_m = \gamma_m - \omega_c\Delta\tau_m + \theta_d -\omega_c\ell T_s$;
    \item $\xi^{(m)}_{n}=\theta_r((n-\ell+(m-1)N)T_s) - \theta_r((n+(m-1)N)T_s)$;
    \item $b^{(m)}_{n}= b((n+(m-1)N)T_s)$.
\end{itemize}
Here, $\phi_m := \gamma_m - \omega_c\Delta\tau_m$ represents the total dynamic phase change induced by the strain, and thus the quantity we actually wish to track. Since $\theta_d$ and $\omega_c\ell T_s$ are constants, we track the change of $\phi_1,\ldots, \phi_M$ through the equivalent changes in the sequence $\psi_1,\ldots, \psi_M$.

The laser phase noise $\theta_r(t)$ is modeled as a Wiener process\cite{modeling}, so that $\xi^{(m)}_{n}$ becomes:
    \begin{equation}
    \xi^{(m)}_{n} = -\sum_{k=n}^{n+\ell-1} \Delta\theta_k.
\label{eq:phase_noise_model}
    \end{equation}
  for the i.i.d. $\Delta\theta_k \sim \mathcal{N}(0, \sigma_{\text{pn}}^2)$ and $\sigma_{\text{pn}}^2=2\pi \Delta\nu T_s$ where $\Delta\nu$ is the linewidth of the laser. The phase noise  $\xi_{n}^{(m)}$ is thus of variance $\ell \sigma_{\textnormal{pn}}^2$, which increases with the delay parameter $\ell$. The thermal noise is modeled as 
a complex circular AWGN with two-sided power spectral density  $N_0$, and finally we assume that the symbols $s_1,\ldots, s_N$ satisfy $|s_n|^2=1$ for all $n=1,\ldots, N$.  For example, Perfect-Squares Minimum-Phase Constant-Amplitude Zero-AutoCorrelation (PS-MP CAZAC) sequences are a common choice %$s_1,\ldots, s_N$ 
\cite{Rohrs1992CAZAC}.

\subsection{The Estimation Problem}
In this paper, we individually estimate each of the phases $\psi_1,\ldots, \psi_M$. This allows to track the change in $\phi_1,\ldots, \phi_M$.

Estimation performance is measured in MSE but accounting for the $2\pi$-periodicity of the phase of a complex number. So, we consider the circular distance
\begin{equation}
\label{eq:circular_distance}
d_c(\psi_m, \hat{\psi}_m) = \vert\mathrm{mod}(\psi_m - \hat{\psi}_m + \pi, 2\pi) - \pi\vert
\end{equation}
where $\mathrm{mod}(x, y)$ denotes the remainder of $x$ modulo $y$, returning a value in $[0, y)$. The $m$-th block estimation performance is given by 
\begin{equation}
\label{eq:cmse_def}
\operatorname{MSE}_{m} =\mathbb{E}\left[ d_c(\psi_m, \hat{\psi}_m)^2 \mid \mathbf{y}_m \right],
\end{equation}
where  $\mathbb{E}[\cdot \mid \cdot]$ denotes  conditional expectation and $\mathbf{y}_m:=[y_{(m-1)N+1}, \ldots, y_{mN}]^T$.

The optimal estimator to minimize \eqref{eq:cmse_def} is  well-known:
\begin{equation}
\label{eq:cmmse_obj}
{\psi}^*_{m,\text{CMMSE}} = \arg\min_{\hat{\psi}_m} \mathbb{E}\left[ d_c(\psi_m, \hat{\psi}_m)^2 \mid \mathbf{y}_m \right],
\end{equation}
but difficult to evaluate directly. One of the  difficulties stems from the circular distance function in  \eqref{eq:circular_distance}. In the following, we therefore approximate both the distance function as well as the signal model and then propose to use the linear MMSE estimator for the approximated model.

%Our goal is to minimize In this way, by applying the estimators for every block, we can achieve the global minimum MSE :
%\begin{equation}
%\label{eq:global_cmse_def}
%\operatorname{MSE}_{\text{global}} =\sum_{m=1}^M \mathbb{E}\left[ d_c(\psi_m, \hat{\psi}_m)^2 \mid \mathbf{y}_m \right],
%\end{equation}

%Therefore, we can optimize the estimation performance for one block and then promote to others. In the following analysis, we only consider one of the blocks, hence the block number $m$ is omitted.

\section{Proposed Phase Estimation Method}
\label{sec:phase_estimation_methods}

We present an estimator $\hat{\psi}_m$ to estimate the block-$m$ phase $\psi_m$ based on the block-$m$ signal vector $\mathbf{y}_m$ with a small MSE as defined in \eqref{eq:cmse_def}. As previously mentioned, finding  the optimal estimator seems intractable both analytically and numerically, and instead we propose a suboptimal estimator based on an approximate loss function and an approximate signal model. %The approximate loss function is described in Subsection~\ref{subsec:cmmse}, the approximate signal model in Subsection~\ref{subsubsec:linearization}, and our proposed estimator in Subsection~\ref{subsubsec:lmmse_derivation}. 
%Before presenting our estimator, we  linearize our loss function in \eqref{eq:cmse_def} and the signal model in \eqref{eq:discrete_signal}. 
We treat all blocks in the same way, and therefore 
%These approximations} apply in the same way to all blocks, and for readability we therefore 
drop the block-index $m$ for readability in the  following.

\vspace*{-10pt}
\subsection{Approximation of the Loss Function}
\label{subsec:cmmse}
We introduce a new  measure
\begin{equation}
    \tilde{\mu}(\psi, \hat{\psi}) := 2 - 2\cos(\psi - \hat{\psi}),
    \end{equation}
    and notice that when $\psi$ and $\hat{\psi}$ are close (modulo $2\pi$): 
    \begin{equation} 
(d_c(\psi, \hat{\psi}))^2 \approx \tilde{\mu}(\psi, \hat{\psi}).\label{eq:max}
    \end{equation}
We thus aim at finding the  optimal estimator for the new estimation problem\begin{IEEEeqnarray}{rCl}
\label{eq:max_cosine}
\psi_{\text{A}}^*&:=& \arg\min_{\hat{\psi}} \mathbb{E}\left[ \tilde{\mu}(\psi ,\hat{\psi}) \mid \mathbf{y} \right] \IEEEyessubnumber* \\
&=&\arg\max_{\hat{\psi}} \mathbb{E}\left[ \cos(\psi - \hat{\psi}) \mid \mathbf{y} \right] 
\end{IEEEeqnarray}
 %  It shows that minimizing \eqref{eq:cmmse_obj} is approximate to maximizing the expected cosine:
%\begin{equation}
%\label{eq:max_cosine}
%\hat{\psi}_{\text{CMMSE}}\approx\arg\max_{\hat{\psi}} \mathbb{E}\left[ \cos(\psi - \hat{\psi}) \mid \mathbf{y} \right] = \hat{\psi}_{\text{ACMMSE}}.
%\end{equation}

Using Euler's formula and the linearity of expectation, and denoting the real part of a complex value $c$ by $\Re(c)$, we can write:
\begin{subequations}
\begin{IEEEeqnarray}{rCl}
\mathbb{E}[\cos(\psi - \hat{\psi}) | \mathbf{y}] &=& \mathbb{E}[\Re(e^{j(\psi - \hat{\psi})}) | \mathbf{y}] \\
&=& \Re(e^{-j\hat{\psi}} \mathbb{E}[e^{j\psi} | \mathbf{y}]) \\
\label{eq:cosine_expectation}
&=& |\mathbb{E}[e^{j\psi} | \mathbf{y}]| \cos(\angle(\mathbb{E}[e^{j\psi} | \mathbf{y}]) - \hat{\psi}),
\IEEEeqnarraynumspace
\end{IEEEeqnarray}
\end{subequations}
where we use $\angle$ to denote the phase of a complex number.
The term above is maximized when the argument of the cosine is $0$ (modulo $2\pi$) and thus 
\begin{subequations}\label{eq:cmmse_solution}
\begin{align} 
{\psi}_{\text{A}}^* &= \angle\left( \mathbb{E}[ e^{j\psi} \mid \mathbf{y} ] \right) = \angle\left( \int_{-\pi}^{\pi} e^{j\psi}  p(\psi \mid \mathbf{y})  d\psi \right)\\
&= \angle\left(\frac{ \int_{\psi} \int_{ \boldsymbol{\xi}} e^{j\psi}p(\boldsymbol{\xi})  p(\mathbf{y} | \psi, \boldsymbol{\xi})  d\boldsymbol{\xi}d\psi }{ \int_{\psi}\int_{\boldsymbol{\xi}} p(\boldsymbol{\xi}) p(\mathbf{y} | \psi, \boldsymbol{\xi}) d\boldsymbol{\xi}d\psi }\right),
\end{align}
\end{subequations}
where the conditional likelihood $p(\mathbf{y} | \psi, \boldsymbol{\xi})$ is a product of independent complex Gaussian densities given a specific phase noise vector $\boldsymbol{\xi}$:
\begin{equation}
\label{eq:prob_ypsixi}
    p(\mathbf{y} | \psi, \boldsymbol{\xi}) = \prod_{n=1}^{N} \frac{1}{2\pi N_0} \exp\left( -\frac{1}{2N_0} \left| y_n - A s_n e^{j(\psi + \xi_n)} \right|^2 \right),
\end{equation}
and $p(\boldsymbol{\xi})$ is the prior probability density function of the phase noise vector $\boldsymbol{\xi}=(\xi_1,\ldots, \xi_N
)$.

Accurate evaluations of above integrals are difficult. To obtain a practical estimator, in the next subsection we propose an approximation of the signal model $\mathbf{y}$   that renders the maximization in \eqref{eq:max}, and thus the maximization of \eqref{eq:max_cosine}, tractable also numerically.

\subsection{The Approximate Signal Model}
\label{subsubsec:linearization}
One of the main challenges in our model is the presence both of multiplicative noise and additive noise. %It is difficult to derive a closed-form solution for this model \eqref{eq:discrete_signal} because we need to consider both multiplicative phase noise and additive thermal noise simultaneously.
However, in practical scenarios the phase noise $\xi_n$ is relatively small, for example  in the sense that $\mathbb{E}[\xi_n^2]= \ell\sigma_{\text{pn}}^2 \ll 1$, which motivates us to use the Taylor approximation\footnote{This Taylor expansion is widely used in engineering systems. A related application is described in \cite{Joint_2006_Darryl}.}
\begin{equation}
\label{eq:taylor_expansion}
e^{j\xi_n} = 1 + j\xi_n + \frac{(j\xi_n)^2}{2} + \ldots \approx 1 + j\xi_n.
\end{equation}
    This transforms the phase noise into an additive noise. %Such small-angle approximations are common accross various fields, see e.g., as previously adopted in~\cite{Joint_2006_Darryl}. 
To see that in practical scenarios $\ell \sigma_{\textnormal{pn}}^2\ll 1$, notice that  a typical transmitter in a $\phi$-OTDR interrogator uses a laser of linewidth around $\Delta\nu = 100$~Hz and wavelength $\lambda = 1550$~nm, and the employed symbold duration is $T_s = 10$~ns. For such parameters and a target distance around $D=200$~m we obtain
\begin{equation}\ell  \sigma_{\textnormal{pn}}^2 =  2 \pi \ell\Delta \nu T_s \approx 0.001256,\label{eq:var}
\end{equation}
because $\ell= 2  D  n_{\lambda}  /(T_s c)\approx 200$, where $c$ denotes the speed of light in vacuum and $n_{\lambda}\approx 1.5$ is the refractive index at wavelength $\lambda$. 
Alternatively, motivation for the small-angle approximation  in \eqref{eq:taylor_expansion} also stems from the fact that  the expected absolute error $\mathbb{E}[|e^{j \xi_n} - (1+j \xi_n)|]$ lies below $10^{-3}$ for above practical parameters, and keeping the same laser it lies below $0.1$ whenever $\ell < 32000$, i.e., when $D<32$ km.

Plugging  \eqref{eq:taylor_expansion} into the signal model  \eqref{eq:discrete_signal} yields:
\begin{align}\label{eq:linearized_y}
y_n &\approx \tilde{y}_n:=A s_n e^{j\psi} (1 + j\xi_n) + b_n. 
\end{align}
Multiplying both sides above by the conjugate of the known transmitted symbol $s_n^*$, we obtain that
\begin{align}
z_n &:= y_n s_n^*
\end{align} 
and 
\begin{align}
   \tilde{z}_n&:= A e^{j\psi} + A e^{j\psi} j \xi_n + w_n \approx z_n
\end{align}
where $w_n = b_n s_n^* \sim \mathcal{CN}(0, 2N_0)$ since we have $|s_n|^2 = 1$. 

Defining 
%\begin{IEEEeqnarray}{rCl}
%\eta &:= &A e^{j\psi} %,\\
%v_n & := &  A e^{j\psi} j \xi_n + w_n,
%\end{IEEEeqnarray}
%and 
$\mathbf{\tilde{z}} = [\tilde z_1, \tilde z_2, \ldots, \tilde z_N]^T$, we  write in vector notation:
\begin{equation}
\label{eq:z_vector_model}
\tilde{\mathbf{z}} =  A e^{j\psi} ( \mathbf{1} + j \boldsymbol{\xi}) +  \mathbf{w},
\end{equation}
where  $\mathbf{1}$ denotes an $N \times 1$ vector of ones, $\boldsymbol{\xi}:= [\xi_1, \xi_2, \ldots, \xi_N]^T$ and $\mathbf{w}:= [w_1, w_2, \ldots, w_N]^T$. 

%Notice that in above model \eqref{eq:z_vector_model}, the ``noise-vector" $\mathbf{v}$ depends on the parameter $\eta$. 
%In the following subsection, we propose an estimator based on the approximate signal model \eqref{eq:z_vector_model} and the approximate estimation problem \eqref{eq:max_cosine}.

\subsection{Our Proposed Estimator}
\label{subsubsec:lmmse_derivation}

In this subsection, we describe our practical estimator and explain how it is motivated by the approximate estimation problem and signal model presented in the previous subsections, see \eqref{eq:max_cosine} and  \eqref{eq:z_vector_model}.

We propose to use the estimator: 
\begin{eqnarray}
\label{eq:final_phase_estimate}
\hat{\psi}_{\text{proposed}} (\mathbf{z})= \hspace{5.5cm} \nonumber\\ \angle\left( |A|^2 \mathbf{1}^T \left( |A|^2 \mathbf{1} \mathbf{1}^T +  |A|^2 \mathbf{\Sigma}_{\xi} + 2N_0 \mathbf{I}_N\right)^{-1}\mathbf{z}\right),
\end{eqnarray}
where $(\cdot)^{-1}$ represents the inverse of the matrix, $\mathbf{I}_N$ denotes the $N \times N$ identity matrix, $\mathbf{z}= [z_1,\ldots, z_N]^{T}$ with $z_n=y_n s_n^*$ 
and 
  the row-$n$ column-$m$ element of $\mathbf{\Sigma}_{\xi}$  is given by:
    \begin{equation}
    \label{eq:sigma_xi}
    (\mathbf{\Sigma}_{\xi})_{n,m} = \sigma_{\text{pn}}^2 \cdot \max(0, \ell - |n-m|).
    \end{equation}
%where $\psi_{\text{proposed}} \approx \psi_{\text{A}}^* \approx \psi_{\text{CMMSE}}$, which approximates to the original optimal estimator \eqref{eq:cmmse_obj} through two approximation \eqref{eq:max_cosine} and \eqref{eq:linearized_y}.

%To motivate our choice, notice  The linearized model \eqref{eq:z_vector_model} fits the standard linear Gaussian estimation framework. To adapt to this framework, the objective function \eqref{eq:max_cosine} can be written in another expression: 
Our choice is motivated by the following two observations. First, we can write
%\begin{equation}
%\label{eq:LMMSE_phase_estimator}
%\tilde{\psi}_{\text{LMMSE}} =
%\angle\left( |A|^2 \mathbf{1}^T \left( |A|^2 \mathbf{1} \mathbf{1}^T +  |A|^2 \mathbf{\Sigma}_{\xi} + 2N_0 \mathbf{I}_N\right)^{-1}\mathbf{\tilde{z}}\right),
%\end{equation}
%Indeed, for $\eta= A e^{j \psi}$, we can write:
\begin{subequations}\label{eq:acmmse_obj}
\begin{align}
\psi_{\text{A}}^*&=\mathop{\arg \min}\limits_{\hat{\psi}\in[-\pi,\pi)} \mathbb{E}\left[ 2-2\cos(\psi-\hat{\psi}) \mid  \mathbf{\tilde{z}} \right]\\
&=\mathop{\arg \min}\limits_{\hat{\psi}\in[-\pi,\pi)} \mathbb{E}\left[ \left\vert Ae^{j\psi} - Ae^{j\hat{\psi}}\right\vert^2 \mid  \mathbf{\tilde{z}} \right]\\
&=\angle \left( \mathop{\arg \min}\limits_{\substack{\hat{\eta} \colon |\hat{\eta}|=A,\\ \hat{\eta}\in\mathbb{C}}} \mathbb{E}\left[ \left\vert \eta - \hat{\eta} \right\vert^2 \mid  \mathbf{\tilde{z}}\right]\right),\label{eq:acmmse_obj_3}
\end{align}
\end{subequations}
and thus finding the optimal estimator in this case becomes similar to the problem of finding the optimal MMSE estimator in a linear Gaussian model in 
%which is the optimal estimator for $\psi$ based on $\mathbf{\tilde{z}}$ in (\ref{eq:z_vector_model}).
(\ref{eq:z_vector_model}). Second, the optimal LMMSE estimator for  (\ref{eq:z_vector_model}) is well-known and equals
\cite{Kay1993Fundamentals}:
\begin{IEEEeqnarray}{rCl}
\tilde{\eta} &:=& \mathsf{K}_{\eta \mathbf{\tilde{z}}}  \mathsf{K}^{-1}_{ \mathbf{\tilde{z}}}\mathbf{\tilde{z}} \\
&=& |A|^2 \mathbf{1}^T \left( |A|^2 \mathbf{1} \mathbf{1}^T +  |A|^2 \mathbf{\Sigma}_{\xi} + 2N_0 \mathbf{I}_N\right)^{-1}\mathbf{\tilde{z}}, 
\end{IEEEeqnarray} 
where $\mathsf{K}_{\eta \mathbf{\tilde{z}}}$ is the cross-correlation matrix between $\eta$ and $\mathbf{\tilde{z}}$ and $\mathsf{K}_{ \mathbf{\tilde{z}}}$ is the auto-correlation matrix of $\mathbf{\tilde{z}}$.
%It minimize $\mathbb{E}\left[ \left\vert \eta - \hat{\eta} \right\vert^2 \mid  \mathbf{\tilde{z}}\right]$ in (\ref{eq:acmmse_obj_3}).% (This is shown in Appendix~A.) 

%In this form, our goal is to find the optimal linear estimator for $\hat{\eta}$ given the received signal $\mathbf{y}=[y_1,y_2,\ldots,y_N]$, satisfying the model \eqref{eq:discrete_signal}.

\section{Cram\'er-Rao Lower Bound}
\label{sec:low_bound}
This section presents the CRLB on minimum mean square error. Notice  that our proposed estimator  in \eqref{eq:final_phase_estimate} is  unbiased due to the circular symmetry of the phase noise and AWGN. Specifically,  notice:
\begin{IEEEeqnarray}{rCl}
    \mathbb{E}[\hat{\psi}] = \mathbb{E}[\angle(\hat{\eta})] = \mathbb{E}[\angle(e^{j\psi}\hat{\eta} e^{-j\psi})] = \psi + \mathbb{E}[\angle(\hat{\eta} e^{-j\psi})].
    \IEEEeqnarraynumspace
\end{IEEEeqnarray} 
Moreover, due to the circular symmetry of the phase noise $\boldsymbol{\xi}$ and the AWGN $\boldsymbol{b}$, the random angle $\angle(\hat{\eta} e^{-j\psi})$ is symmetric around zero, and hence  $\mathbb{E}[\angle(\hat{\eta} e^{-j\psi})]=0$,  establishing the  unbiasedness of the estimator. As a consequence, the CRLB provides a lower bound on its MMSE. Our numerical results indicate that it also lower bounds the expected square loss under the circular loss function in \eqref{eq:circular_distance}.
%we observe that the Cramer-Rao lower bound  and  the derives the theoretical and numerical low bounds for phase estimation under the joint impairment of thermal noise and the phase noise. 

%We first derive the theoretical CRLB for  our original signal model in \eqref{eq:discrete_signal}.

%But we derive two asymptotic bounds for the CRLB. Then, we use a Monte Carlo method to compute CRLB approximately.  Finally, we analyze the behavior of the numerical CRLB in different SNR regimes and present the expression of low bound in practice.

\subsection{The CRLB}

CRLB states that for any unbiased estimator $\hat{\psi}$\cite{Bay2008Analytic}:
\begin{equation}
  \operatorname{Var}(\hat{\psi}) \geq \operatorname{CRLB}(\psi) = \frac{1}{I(\psi)},\label{eq:CRLB}
\end{equation}
where  $I(\psi)$ denotes the Fisher information 
\begin{equation}\label{eq:fisher_info}
    I(\psi) = \mathbb{E}_{\mathbf{y} | \psi} \left[ \left( \frac{\partial \log p(\mathbf{y} | \psi)}{\partial \psi} \right)^2 \right]
\end{equation}
and 
$p(\mathbf{y} | \psi)$ is  characterized by our signal model in \eqref{eq:discrete_signal}.
% the marginal likelihood $p(\mathbf{y} | \psi)$ is obtained by integrating out the laser phase noise vector $\boldsymbol{\xi}$:
%\begin{equation}
%\label{eq:prob_ypsi}
%    p(\mathbf{y} | \psi) = \int p(\boldsymbol{\xi})  p(\mathbf{y} | \psi, \boldsymbol{\xi})  \mathrm{d}\boldsymbol{\xi},
%\end{equation}
%where $p(\boldsymbol{\xi})$ follows a zero-mean multivariate Gaussian distribution with covariance matrix $\boldsymbol{\Sigma}_{\xi}$.
%The conditional log-likelihood is:
In fact, based  on this signal model, we can write:
\begin{IEEEeqnarray}{rCl}
    \frac{\partial \log p(\mathbf{y} | \psi)}{\partial \psi} &=& \frac{ \partial  p(\mathbf{y} | \psi)}{\partial \psi} \cdot \frac{1}{ p(\mathbf{y} | \psi)}\\ &=& \frac{ \int p(\boldsymbol{\xi})  p(\mathbf{y} | \psi, \boldsymbol{\xi})  g(\mathbf{y}, \psi, \boldsymbol{\xi})  \mathrm{d}\boldsymbol{\xi} }{ \int p(\boldsymbol{\xi})  p(\mathbf{y} | \psi, \boldsymbol{\xi})  \mathrm{d}\boldsymbol{\xi} },\label{eq:dev_cond_log_prob}
\end{IEEEeqnarray}
where $p(\mathbf{y} | \psi, \boldsymbol{\xi})$ is given in \eqref{eq:prob_ypsixi} and %$g(\psi, \boldsymbol{\xi})$ %is the gradient %of the conditional log-likelihood. From the expression of $p(\mathbf{y} | \psi, \boldsymbol{\xi})$, we derive:
\begin{subequations}
\label{eq:gradient_calculation}
    \begin{align}
         g(\mathbf{y}, \psi, \boldsymbol{\xi}) &:= \frac{\partial \log p(\mathbf{y} | \psi, \boldsymbol{\xi})}{\partial \psi} \\
         &= \frac{1}{N_0} \sum_{n=1}^{N} \Im{ \left(y_n^* s_n e^{ j \left( \psi + \xi_n \right) }\right)},   
    \end{align}
\end{subequations}
where $\Im(\cdot)$ represents the imaginary part of the complex value.

Numerical evaluation of the CRLB and $I(\psi)$ is challenging because  expectation in  \eqref{eq:fisher_info} and the integral in \eqref{eq:dev_cond_log_prob} cannot be computed analytically. 
We employ a double Monte Carlo method as follows:
\begin{itemize} 
\item 
\textit{Outer Loop for computing  the expectation in \eqref{eq:fisher_info}:} Randomly pick $\psi$ and generate $K$ phase noise paths $\boldsymbol{\xi}^{(k)}$  and additive noises $\mathbf{w}^{(k)}$, for $k=1,\ldots, K$,  by drawing them independently according to the distributions $p_{\boldsymbol{\xi}}$ and $p_{\boldsymbol{w}}$, and then generate the corresponding observation vectors $\{\mathbf{y}^{(k)}\}_{k=1}^{K}$ according to the true system model \eqref{eq:discrete_signal} and  the chosen phase $\psi_s$. 

The score function is  computed as \begin{equation}
    I(\psi_s) \approx \frac{1}{K} \sum_{k=1}^{K} \left[ \left. \frac{\partial \log p(\mathbf{y}^{(k)} | \psi)}{\partial \psi} \right|_{\psi = \psi_s} \right]^2.
\end{equation}

\item \textit{Inner Loop for computing the integral in \eqref{eq:dev_cond_log_prob}:} For each $k=1,\ldots, K$, generate $P$ independent phase noise paths $\{\boldsymbol{\xi}^{(p)}\}_{p=1}^{P}$ according to $p_{\boldsymbol{\xi}}$ and use the approximation:
\begin{equation}
    \label{eq:score_calculation}
    \frac{\partial \log p(\mathbf{y}^{(k)} | \psi)}{\partial \psi}
    \approx \frac{ \sum_{p=1}^{P} p(\mathbf{y}^{(k)} | \psi, \boldsymbol{\xi}^{(p)}) g(\psi, \boldsymbol{\xi}^{(p)}) }{ \sum_{p=1}^{P} p(\mathbf{y}^{(k)} | \psi, \boldsymbol{\xi}^{(p)}) },
\end{equation}
where the gradient $g(\psi, \boldsymbol{\xi}^{(p)})$ is computed via \eqref{eq:gradient_calculation}. %We set $w_{kp}=p(\mathbf{y}^{(k)} | \psi, \boldsymbol{\xi}^{(p)})$ as the weight of this gradient, then \eqref{eq:score_calculation} can be expressed as:
%\begin{equation}
    %\label{eq:weght_score_calculation}
  %  \frac{\partial \log p(\mathbf{y}^{(k)} | \psi)}{\partial \psi}
 %   \approx \frac{\sum_{p=1}^{P}w_{kp} g(\psi, \boldsymbol{\xi}^{(p)}) }{ \sum_{p=1}^{P} w_{kp} },
%\end{equation}
%where the choose of path $\{\boldsymbol{\xi}^{(p)}\}_{p=1}^{P}$ determine the approximate accuracy.
\end{itemize}

Numerical simulations indicate that the Fisher information $I(\psi)$ is equal for all  values of $\psi \in[-\pi,\pi)$,  which therefore represents a lower bound on the MMSE. We denote this value as $\operatorname{CRLB}_{\text{num}}$ in the following. 

\subsection{Analytic CRLBs for Idealized Models}
Omitting either the phase noise $\xi_n$ or the AWGN $w_n$ from the discrete-time model \eqref{eq:discrete_signal} allows for improved estimation. The CRLBs for these simplified models lower bounds the original CRLB in \eqref{eq:CRLB}  and  can be calculated analytically. 
\subsubsection{No AWGN---only Phase Noise}
The CRLB bound for the model in \eqref{eq:discrete_signal} without AWGN (i.e., specialized to  $w_n=0$ deterministically) can be calculated  in closed form as:
\begin{equation}
    \operatorname{CRLB}_{\text{PN}} = \frac{1}{\mathbf{1}^T \boldsymbol{\Sigma}_{\xi}^{-1} \mathbf{1}},
\end{equation}
Notice that this bound does not depend on the amplitude $A$, and is thus independent of the SNR:=$|A|^2/N_0$.  This bound is expected to approach the original CRLB bound in the high SNR regime where the influence of the AWGN becomes negligible compared to the phase noise.

\subsubsection{No Phase Noise---only AWGN}
The CRLB bound for the model \eqref{eq:discrete_signal} without phase noise  (i.e.,  $\xi_n=0$ deterministically) evaluates to:%$\operatorname{SNR} \to -\infty$, this problem become the classical linear phase estimation \cite{Kay1993Fundamentals}. Its CRLB is:
\begin{equation}
    \operatorname{CRLB}_{\text{AWGN}} = \frac{N_0}{\sum_{n=1}^{N} |A s_n|^2}
\end{equation}
%This bound decreases inversely with the SNR.
Notice that it does not depend on the delay parameter $\ell$. It is expected to be close to the CRLB bound at low SNR regime, because there the influence of the AWGN seems dominant compared to the phase noise. 

\begin{figure}[t!]
	\centering
	\includegraphics[width=0.45\textwidth]{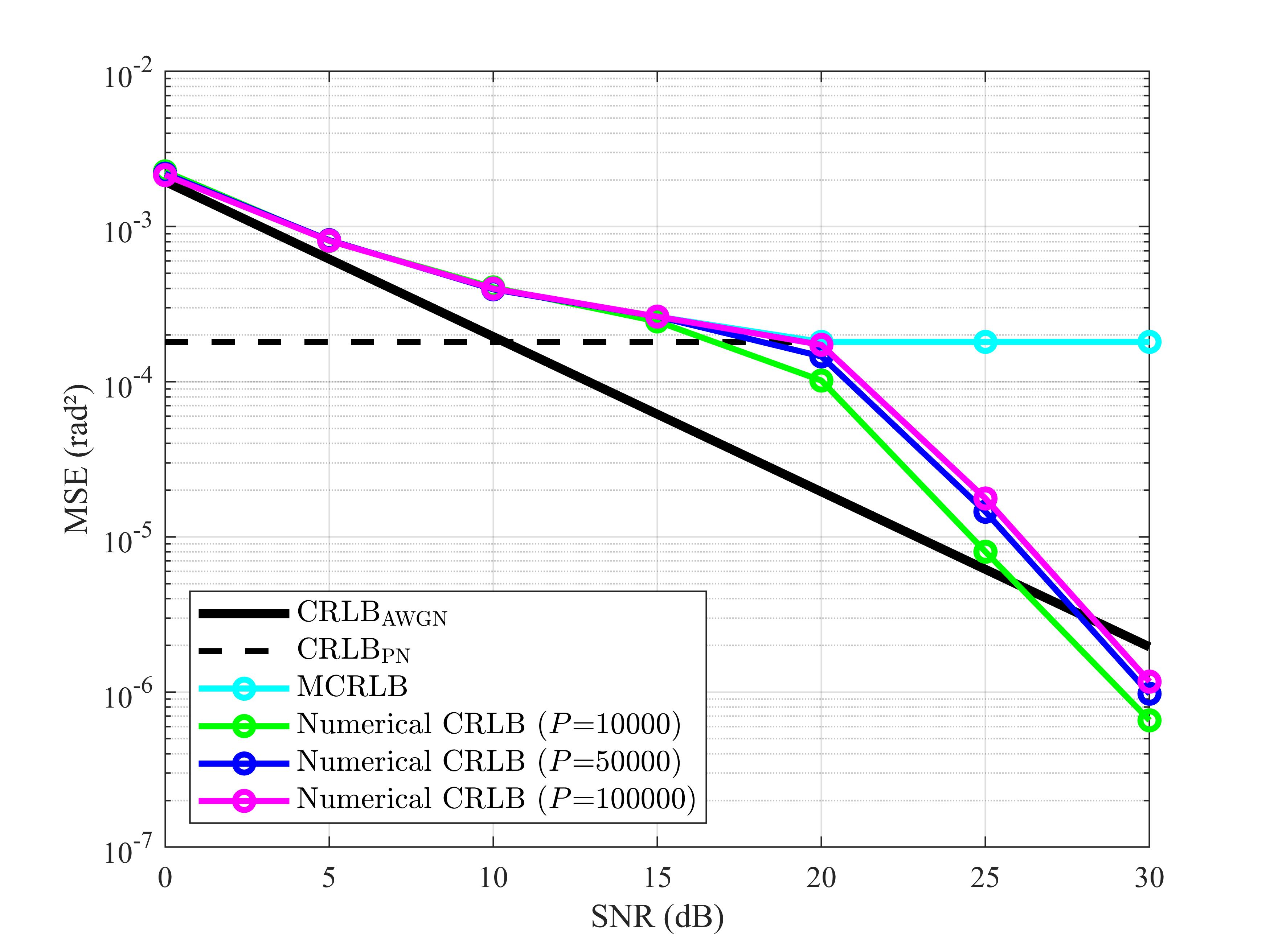}
	\caption{Performance comparison of the numerical CRLBs with different paths and the asymptotes for CRLB when $\ell=100$.}
	\label{fig:num_CRLBs}
\end{figure}

\begin{table}[b!]
\vspace{-7mm}

\centering
\caption{SIMULATION PARAMETERS}
\label{tab:simulation_parameters}
\begin{tabular}{ c c c }
\toprule
\textbf{Notation} & \textbf{Definition} & \textbf{Value} \\ 
\midrule
$N$ & Sequence Length & 256 \\
$\ell$ & Delay Index & $\{0,10,100, 200\}$ \\ 
$\Delta\nu$ & Linewidth & 100 Hz \\ 
$T_s$ & Symbol Duration & 10 ns \\ 
$\sigma_{\text{pn}}^2$ & Phase Noise Variance & $6.2832\times10^{-6}$ \\ 
\bottomrule
\end{tabular}
\end{table}
\subsection{Numerical Evaluation of the CRLBs}
Fig.~\ref{fig:num_CRLBs} shows the  numerical evaluation of the described double Monte Carlo method for computing the CRLB lower bound $\operatorname{CRLB}_{\text{num}}$. Simulation parameters are as given in Table~\ref{tab:simulation_parameters} while picking $K=5000$, $P=10 000$, $50 000$, and $100 000$ for the double Monte Carlo loops. Under the chosen parameter values, the different delay parameters $\ell \in\{0,10,100, 200\}$ correspond to  FBG distances $D \in\{0,10, 100, 200\}$ m.

%Henceforth, we denote this curve by  
We also plot the simpler analytic CRLB bounds without AWGN and without phase noise. As mentioned, they should provide lower bounds on $\operatorname{CRLB}_{\text{num}}$ and be tight at high- and low-SNR respectively.  We observe that this is indeed the case at  SNRs up to 15 dB, while afterwards,  $\operatorname{CRLB}_{\text{num}}$ lies below $\operatorname{CRLB}_{\text{PN}}$, indicating that our numerical evaluation of  CRLB$(\psi)$ is inaccurate and underestimates the actual value. 

% This phenomenon seems to be caused by the Monte Carlo evaluation of the integral in the inner loop of our numerical evaluation algorithm. In fact, while at  low SNRs the conditional likelihood { \color{red}$p_{\mathbf{y}^{(k)}|\psi, \boldsymbol{\xi}}$ (change into $f(\boldsymbol{\xi})$)} is relatively flat over all realizations of the phase noise $\boldsymbol{\xi}$, at high SNRs it becomes very sharp.  Since we are randomly sampling the realizations (paths) $\boldsymbol{\xi}^{(p)}$ that we use to approximate the integral,  there is a significant chance that we miss the peak and the integral will be overestimated, resulting in a underestimation of the CRLB bound.

This phenomenon is caused by the Monte Carlo evaluation of the right-hand-side term in \eqref{eq:score_calculation} using a limited number of phase noise realizations $\boldsymbol{\xi}$. This is problematic at high SNR values. At low SNR, the conditional likelihood $p(\mathbf{y}^{(k)}|\psi,\boldsymbol{\xi})$ is relatively flat over $\boldsymbol{\xi}$, so 10000 randomly generated paths $\boldsymbol{\xi}^{(p)}$ from the prior $p(\boldsymbol{\xi})$ are enough for a good approximation of~\eqref{eq:score_calculation}. However, as SNR increases, $p(\mathbf{y}^{(k)}|\psi,\boldsymbol{\xi})$  becomes  sharper around the true phase noise, and given that the prior distribution $p(\boldsymbol{\xi})$ does not depend on SNR, most of the sampled $\boldsymbol{\xi}^{(p)}$ miss the peak even when we generate 100000 paths for $\text{SNR}>20$~dB. Consequently, the Monte Carlo method fails to provide a good approximation of~\eqref{eq:score_calculation}, leading to distorted numerical results for the CRLB at high SNR. %Simulation results show that the sum $\sum_{p=1}^{P} p(\mathbf{y}^{(k)}|\psi,\boldsymbol{\xi}^{(p)})\ll 1$ in the denominator of (\ref{eq:score_calculation}), which indicates that the sampled paths fail to capture the region where the likelihood is significant. Consequently, the Monte Carlo method fails to provide a good approximation of the integral, leading to distorted numerical results for the CRLB.}

To circumvent the above problem,  in our numerical simulations,  we  consider the lower bound
\begin{equation}
\label{eq:MCRLB}
    \operatorname{MCRLB} := \max\left( \operatorname{CRLB}_{\text{num}}, \operatorname{CRLB}_{\text{PN}} \right).
\end{equation}

\section{Numerical Results}

\begin{figure}[t!]
	\centering
	\includegraphics[width=0.5\textwidth]{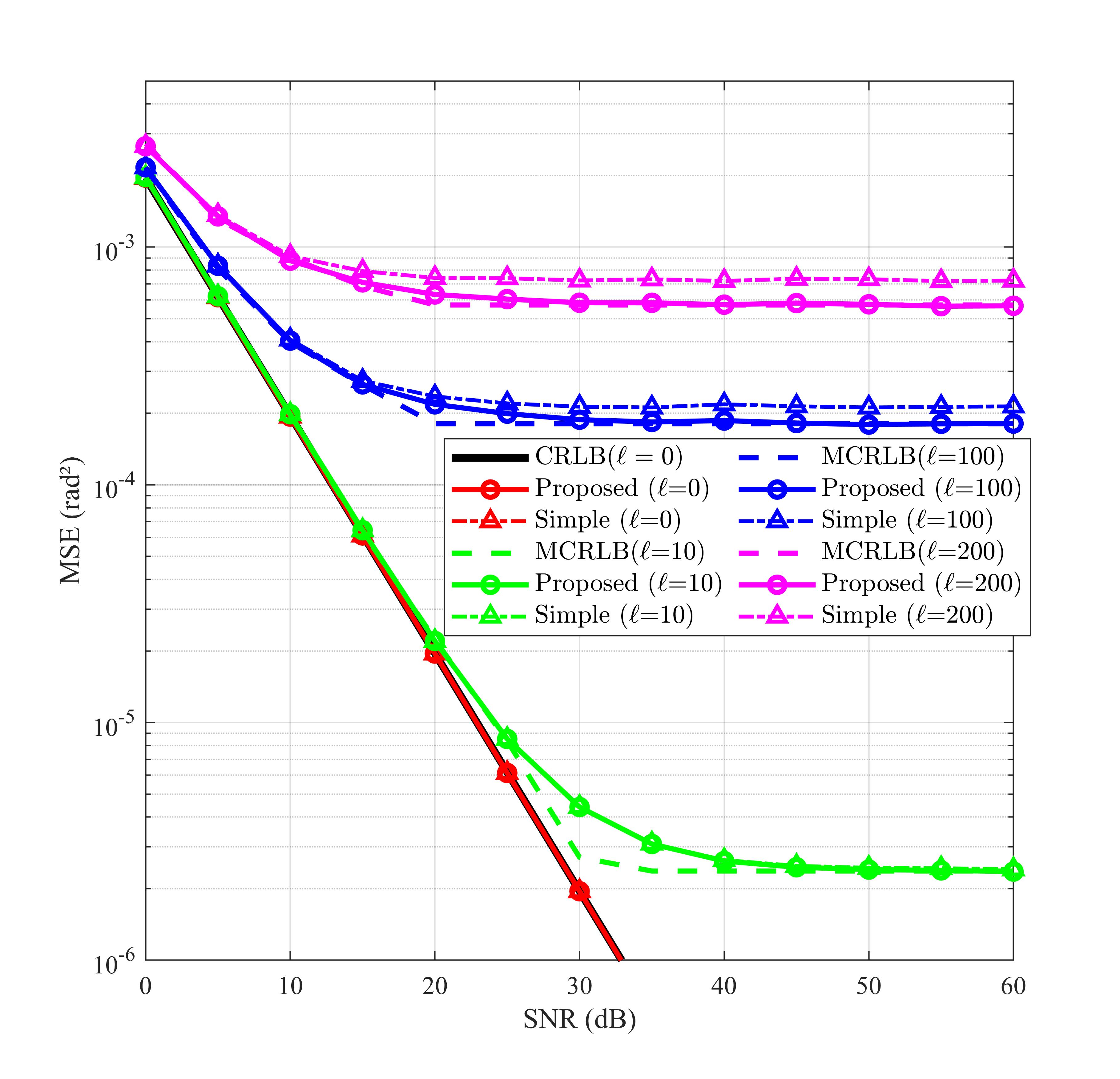}
	\caption{Performance comparison of the estimators against the asymptotes.}
	\label{ammse}
\end{figure}

We numerically evaluate the proposed estimator  in \eqref{eq:final_phase_estimate}  and compare it to  the  MCRLB in \eqref{eq:MCRLB} and a simple estimator commonly used in practice, given by 
\begin{equation}
\label{eq:simple_estimator}
\hat{\psi}_{\text{simple}}(\mathbf{z})=\angle{\left(\sum_{n=1}^N z_n\right)},
\end{equation}
Notice that this simple estimator is optimal when the phase noise  $\xi_1,\ldots, \xi_N$ is not present. We thus expect it to have good performance at low signal-to-noise ratios.

%\subsection{Parameter Setup}
Numerical results are based on the spread-spectrum FBG-based sensing systems with  parameters as given in Table~\ref{tab:simulation_parameters}. Fig.~\ref{ammse} shows the MSE $\mathbb{E}\left[ d_c(\psi, \hat{\psi})^2 \right]$\footnote{Notice that even though we derive our estimator $\hat{\psi}_{\text{proposed}}$ in \eqref{eq:final_phase_estimate} based on the approximate signal model in \eqref{eq:linearized_y} and the approximate loss function \eqref{eq:max}, our numerical results evaluate $\hat{\psi}_{\text{proposed}}$ on the actual model \eqref{eq:discrete_signal} and the actual loss $\mathbb{E}\left[ d_c(\psi, \hat{\psi})^2 \right]$} achieved by our proposed estimator $\hat{\psi}_{\text{proposed}}$ and the simple estimator $\hat{\psi}_{\text{simple}}$, where expectations are calculated using the Monte Carlo method (50000 $\mathbf{z}$-samples). To obtain the points on the  CRLB bounds, we apply  $K=5000$ outer loops and $P=100000$ inter loops for the Monte Carlo approximation.

\begin{figure}[t!]
	\centering
	\includegraphics[width=0.35\textwidth]{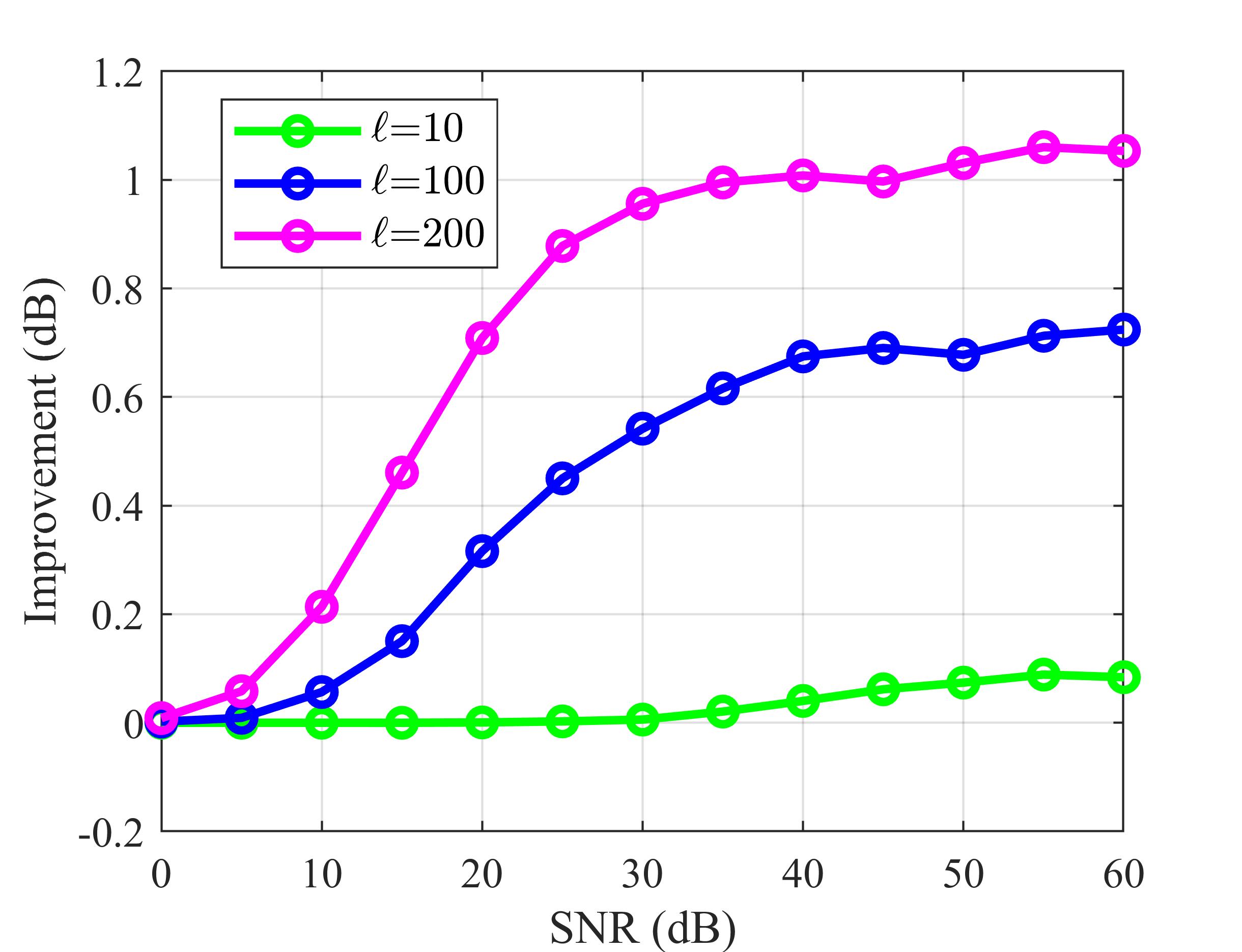}
	\caption{Performance improvement versus SNR for different $\ell$. (Parameters are as in Table~\ref{tab:simulation_parameters}.)}
	\label{fig:gain}
\end{figure}

We observe that at low SNR values, both estimators $\hat{\psi}_{\text{simple}}$ and $\hat{\psi}_{\text{proposed}}$ achieve performances close to the MCRLB. For larger SNR values, our proposed estimator outperforms the simple estimator and manages to achieve performances close to the MCRLB for most SNR values. In particular, it achieves the same MSE  error floor as MCRLB while the simple estimator saturates at a larger MSE value.  The reason is that the phase noise becomes dominant at high SNRs compared to the additive Gaussian noise, and the simple estimator fails to compensate for the phase noise.

Observe also that larger values of $\ell$, i.e., larger delays, cause a higher and earlier error floor compared to  smaller values of  $\ell$. For all tested values of $\ell$, our estimator performs close to the MCRLB bound over the entire SNR range and saturates at the MCRLB error floor.

To further compare the performance of  $\hat{\psi}_{\text{simple}}$ and $\hat{\psi}_{\text{proposed}}$, we plot in Fig.~\ref{fig:gain}, for different values of $\ell$, the ratio 
\begin{equation}
    \text{G (dB)} := 10\log_{10}\left(\frac{\operatorname{MSE}_{\text{simple}}}{\operatorname{MSE}_{\text{proposed}}}\right)
    \label{eq:improvement_ratio}
\end{equation}
where $\operatorname{MSE}_{\text{simple}}$ and $\operatorname{MSE}_{\text{proposed}}$ denote the MSE of the simple estimator and the proposed LMMSE estimator, respectively. We observe that the improvement $\text{G}$ increases with the delay parameter $\ell$ and the SNR, because the phase noise is more harmful in these regimes and better mitigated by $\hat{\psi}_{\text{proposed}}$.

Observe also that larger $\ell$  yields a larger gain of the proposed estimator over the simple estimator. For example, for SNR $=25$ dB, the gain increases from $0.45$ to $0.88$ dB, %$0.719$~dB for $\ell = 100$ to $1.072$~dB for $\ell = 200$, 
indicating that the proposed estimator is more advantageous in longer-delay systems where phase noise is more severe.

\begin{figure}[t!]
	\centering
	\includegraphics[width=0.45\textwidth]{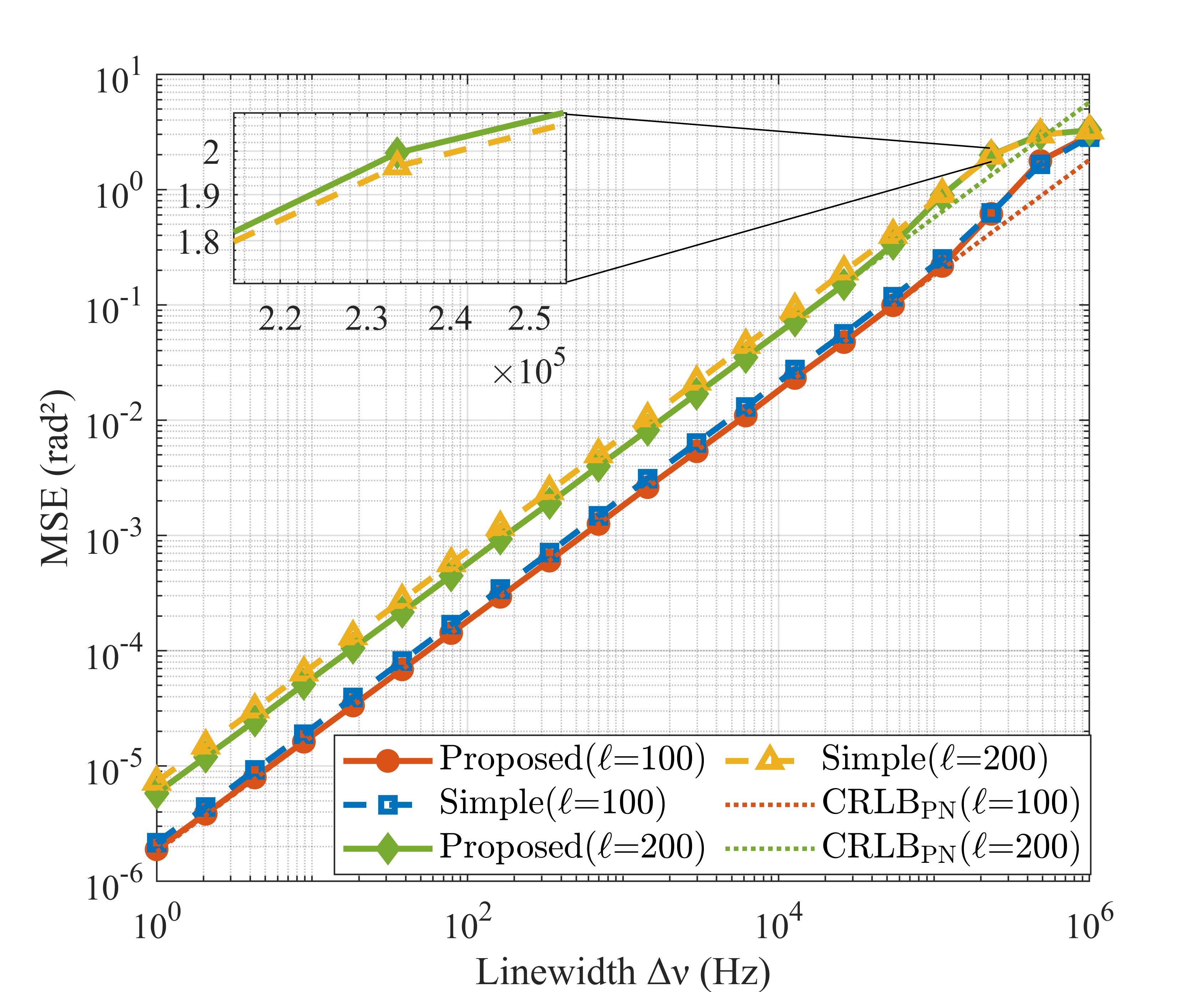}
	\caption{MSE versus laser linewidth $\Delta\nu$ for different $\ell$ when SNR $=50$~dB.}
	\label{fig:mse_linewidth}
\end{figure}

Finally, we evaluate the effect of the laser linewidth on the MSE via Figure~\ref{fig:mse_linewidth}, under otherwise identical system parameters as given in Table \ref{tab:simulation_parameters}. Observe that for  laser linewidth $\Delta\nu < 34$~kHz, the proposed estimator  achieves a gain of approximately $0.8$~dB over the simple estimator and remains close to the CRLB$_{\text{PN}}$. However, for $\Delta\nu > 120$kHz, the proposed estimator performs no better, or even worse, than the simple estimator. This degradation is attributed to the failure of the small-angle approximation $e^{j\xi_n} \approx 1 + j\xi_n$, which becomes invalid when $\xi_n$ grows sufficiently large. More generally, under the considered system parameters, the approximation seems valid for laser linewidth approximately below $1.2 \times 10^5$~Hz, and our proposed estimator performs well in this regime but fails afterwards. To put this in context, one  has to notice that the regime of failure is practically irrelevant because in this regime the MSE is in the order of $1$ and thus already too large for any practical purposes. %\texttt{Influence of $\ell$??}

These numerical results show that the performance gain of the proposed estimator is fundamentally governed by the product $\ell\sigma_{\text{pn}}^2$. As shown in Fig.~\ref{fig:gain} and Fig.~\ref{fig:mse_linewidth}, a larger sensing distance $\ell$ or a larger phase-noise variance $\sigma_{\text{pn}}^2$ both lead to a more pronounced improvement over the standard estimator, due to the increasing impact of the accumulated phase noise.

Our numerical  results thus show that in practically-relevant scenarios our proposed estimator can accurately estimate the phase noise  in fiber-optic spread-spectrum FBG-based sensing systems and improves over the simple standard estimator. 

\vspace*{-5pt}
\section{Conclusion}
This paper establishes a signal model for fiber-optic spread-spectrum FBG-based sensing systems with a single reflector, and proposes a practical estimator to estimate the phase changes in the system. Furthermore, a fundamental CRLB is presented including a numerical algorithm for its evaluation.

Numerical simulation results demonstrate that the proposed estimator successfully compensates for the  phase noise in the FBG-based sensing system and its MSE closely approaches the CRLB, confirming its effectiveness and near-optimality. %We provides a practical method for compensating the phase noise and enhancing the performance of the spread-spectrum DAS systems, enabling high-distance phase tracking under complex environment for various applications.

Interesting future research directions involve testing our  estimator $\hat{\Psi}_{\textnormal{proposed}}$ on a real FBG sensing testbed or to extend our results to  distributed acoustic sensing (DAS), which is able to simultaneously track phase changes in various parts of the fiber. In a DAS model,  multiple Rayleigh scatterers are present, and inter-segment interference arises due to the superposition of backscattered signals with different delays, phases and amplitudes. inter-segment interference effectively degrades estimation performance and renders the task of finding an optimal or near-optimal estimator much more complex. %Technically speaking, The main challenges 
include suppression of inter-segment interference, phase noise compensation across segments, mitigation of polarization effects, and design of probing sequences. Addressing these challenges will be the focus of  future work.

\section*{Acknowledgments}
 Y. Bai acknowledges funding support from the China Scholarship Council, E. Awwad from the ANR under grant agreement ANR-24-CE42-3114 and  M. Wigger from the ERC under grant agreement  101125691.

\bibliographystyle{IEEEtran}
\bibliography{stateOfArt}
\end{document}